\documentclass{ws-ijgmmp}

\usepackage{amsmath}

\begin{document}

\title{A note on the dynamical system formulations in $f(R)$ gravity}

\author{SAIKAT CHAKRABORTY}

\address{Center for Gravitation and Cosmology, Yangzhou University, 88 South University Ave., Yangzhou 225009, Jiangsu, China\\
\email{saikatnilch@gmail.com,\,snilch@yzu.edu.cn}}

\author{PETER K.S. DUNSBY}

\address{Department of Mathematics and Applied Mathematics, Cosmology and Gravity Group, University of Cape Town, Rondebosch 7701, Cape Town, South Africa\\
\email{peter.dunsby@uct.ac.za}}

\author{KELLY MACDEVETTE}

\address{Department of Mathematics and Applied Mathematics, Cosmology and Gravity Group, University of Cape Town, Rondebosch 7701, Cape Town, South Africa\\
\email{kellymacdev@gmail.com,\, MCDKEL004@myuct.ac.za}}

\maketitle

\begin{abstract}
A number of dynamical system formulations have been proposed over the last few years to analyse cosmological solutions in $f(R)$ gravity. The aim of this article is to provide a brief introduction to the different approaches, presenting them in a chronological order as they appeared in the history of the relevant scientific literature. In this way we illuminate how the shortcoming(s) of an existing formulation encouraged the development of an alternative formulation. Whenever possible, a 2-dimensional phase portrait is given for a better visual representation of the dynamics of phase space. We also touch upon how cosmological perturbations can be analyzed using the phase space language. 
\end{abstract}


\newpage

\tableofcontents

\section{Introduction}
$f(R)$ gravity is an important and well-studied class of modified gravity theories \cite{Sotiriou:2008rp,DeFelice:2010aj}. Many important cosmological models, for example Starobinsky's inflationary models \cite{Starobinsky:1980te} and Hu-Sawicki's late time cosmology model \cite{Hu:2007nk}, employ these theories to explain the late-time acceleration of the universe, without the need for dark energy. The field equations for $f(R)$ gravity are comprised of a modified Friedmann equation, a modified Raychaudhuri equation and the continuity equation for the cosmological fluid. In general they constitute a complicated set of coupled non-linear partial differential equations. In cosmology we mostly work with homogeneous metrics in order to respect the cosmological principle, so that the cosmological field equations reduce to a set of coupled non-linear ordinary differential equations. As we know, there are no generic method to solve non-linear differential equations, so  the method of dynamical system can be a good approach to extract qualitative and qualitative information about the evolution of such non-linear systems. In the dynamical system formulation of cosmology, a set of dimensionless expansion normalised dynamical variables are suitably defined in order that the cosmological field equations can be written in as a closed set of coupled first order nonlinear differential equations, constituting a dynamical system \cite{ellisbook,dsa_coley}. The phase trajectories in the corresponding phase space represent all possible cosmological dynamics within the model under consideration. The fixed points are interpreted as cosmic epochs, with unstable fixed points (repelers or past attractors), saddle fixed points and stable fixed points (attractors or future attractors) representing possible initial, intermediate and final epochs of the cosmic evolution.

Dynamical system methods have been extensively used in the context of dark energy models and modified gravity (See Ref.\cite{Bahamonde:2017ize} for a review). For $f(R)$ gravity the very first attempt to employ a phase space analysis can be traced back to Starobinsky \cite{Starobinsky:1980te} in the context of $R^2$-inflation model. Capozziello et al. generalised this study for a number of different inflationary $f(R)$ models \cite{Capozziello:1993xn}. The so called expansion-normalized dynamical variables, which are mostly used today for dynamical system analysis, was introduced for the first time by Carloni et. al. \cite{Carloni:2004kp} for the simple case of $R^n$ gravity. Attempts to generalise the formulation followed shortly afterwards \cite{Amendola:2006we,Carloni:2007br}.  Extensions of this formulation to a compact phase space, allowing for the analysis of asymptotic behaviour including static and cyclic solutions, were also introduced in reference to a number of $f(R)$ models \cite{Abdelwahab:2011dk,Goheer:2009fs}. Despite its success, this formulation is useful only for very certain types of $f(R)$. This shortcoming led Carloni to propose a different dynamical system formulation that can in principle be used for all the $f(R)$ theories \cite{Carloni:2015jla}. A \emph{form-independent} dynamical system approach, that can be used for all possible $f(R)$ theories satisfying an observational requirement irrespective of their functional form, was proposed very recently by Chakraborty et al. \cite{Chakraborty:2021jku}. 

The article is arranged as follows. Sec.\ref{sec:fe} presents a brief introduction to $f(R)$ gravity and the cosmological field equations field equations for $f(R)$ gravity. Sec.\ref{sec:dsa_0} presents the earliest and the very basic dynamical system formulation for $f(R)$ gravity \cite{Starobinsky:1980te,Capozziello:1993xn,deSouza:2007zpn}. Sec.\ref{sec:dsa_1} reviews the usual and most used dynamical system formulation in terms of expansion-normalized dynamical variables \cite{Amendola:2006we,Carloni:2007br} and points out it's shortcoming. Sec.\ref{sec:dsa_2} describes the alternative dynamical system formulation as was proposed by Carloni \cite{Carloni:2015jla}. Sec.\ref{sec:dsa_3} presents the form-independent dynamical system formulation as proposed by Chakraborty et al. \cite{Chakraborty:2021jku}. Sec. \ref{sec:compact} briefly introduces the compact dynamical system formulation and presents some of its strengths and flaws. Finally a discussion is presented in Sec.\ref{sec:disc}. 

Throughout the article we will mainly consider the homogeneous and isotropic Friedmann-L\^{a}imetre-Robertson-Walker (FLRW) spacetime to respect the cosmological principle
\begin{equation}
    ds^2 = -dt^2 + \frac{a^2(t)}{\left(1+\frac{kr^2}{4}\right)^2}\;.
\end{equation}
We will also take into consideration only the metric formulation of $f(R)$ gravity in which the only dynamical degree of freedom is the metric and the affine connection is a function of the metric and it's first derivatives. Another formulation of $f(R)$ gravity is the so-called Palatini formulation, which treats the metric and the affine connection as two independent dynamical quantities. For general relativity (GR) the two formalisms produce the same field equations but for $f(R)$ gravity they produce different field equations. For an interesting connection between these two formulations, see Ref.\cite{Capozziello:2010ef}.
\section{Cosmological field equations in $f(R)$ gravity} \label{sec:fe}
$f(R)$ gravity is characterised by the existence of a dynamical scalar degree of freedom $\varphi=F(R)\equiv f'(R)$ as apparent from the trace field equation 
\begin{equation}\label{tfe}
    RF(R)-2f(R)+3\Box F(R)=\kappa T,
\end{equation}
$T$ being the trace of the energy momentum tensor,  $\kappa=8\pi G=\frac{8\pi}{m_{\rm Pl}^2}$ ($G$ is Newton's gravitational constant and $m_{\rm Pl}$ is the Planck's mass). This propagating scalar degree of freedom is sometimes referred to as the \emph{scalaron}. General Relativity is the trivial case of $f(R)$, for which $F(R)=1$ and the scalaron ceases to be a dynamical degree of freedom. In the presence of a perfect fluid with energy density $\rho$ and pressure $P$, the field equations for FLRW spacetime in $f(R)$ gravity are given by
\begin{eqnarray}
&& 3F\left(H^2 +\frac{k}{a^2}\right)=\kappa\rho_{\rm eff}\equiv\kappa(\rho+\rho_{R})\;,\label{fe1}\\
&& -F\left(2\dot{H}+3H^2 +\frac{k}{a^2}\right)=\kappa P_{\rm eff}\equiv\kappa(P+P_{R})\;,\label{fe2}
\end{eqnarray}
where we have defined the conserved scalaron energy density and pressure as
\begin{eqnarray}
&& \kappa\rho_{R} \equiv \frac{1}{2}(RF-f) - 3H\dot{F}\;,\\
&& \kappa P_{R} \equiv \ddot{F} + 2H\dot{F} - \frac{1}{2}(RF-f)\;.
\end{eqnarray}
The effective equation of state parameter of the universe is defined as \begin{equation}\label{w_eff}
    w_{\rm eff} \equiv \frac{P_{\rm eff}}{\rho_{\rm eff}} = \frac{P + P_R}{\rho + \rho_R} = -\frac{2\dot{H} + 3H^2 + k/a^2}{3(H^2 + k/a^2)}
\end{equation}
and the equation of state of the scalaron is
\begin{equation}\label{w_R}
    w_R \equiv \frac{P_R}{\rho_R} = \frac{\ddot{F} + 2H\dot{F} - \frac{1}{2}(RF-f)}{\frac{1}{2}(RF-f) - 3H\dot{F}}\,\;.
\end{equation}
If the perfect fluid is barotropic, with an equation of state parameter $w$, then $w_{\rm eff}$ and $w_R$ are related to each other via the relation
\begin{equation}\label{www}
    w_{\rm eff} = w\frac{\rho}{\rho_{\rm eff}} + w_R\frac{\rho_R}{\rho_{\rm eff}}\,.
\end{equation}
There are two important conditions for physical viability of any $f(R)$ gravity which we now mention below:
\begin{itemize}
    \item $f'(R)<0$ makes the scalar degree of freedom appearing in the theory a ghost. To eradicate the possibility of a ghost degree of freedom, one must require $f'(R)>0$ for all $R$ (or at least within the domain under consideration). We will assume it to hold throughout our consideration.
    \item $f''(R)<0$ is related to unstable growth of curvature perturbations in the weak gravity limit (This is also known as the tachyonic instability or the Dolgov-Kawasaki instability \cite{Dolgov:2003px}). Therefore, one requires that $f''(R)>0$ at least around the matter domination epoch.
\end{itemize}

In the next few sections we present the various dynamical system formulations for $f(R)$ gravity that exists in literature.
\section{The earliest dynamical system formulation} \label{sec:dsa_0}
The fundamental difference between Einstein's general relativity and $f(R)$ gravity is the existence of terms containing the fourth order derivatives of the metric. This is why $f(R)$ theories are also a subclass of the so-called fourth order theories of gravity. This is apparent from the term $\ddot{F}$ in $P_R$. This basic knowledge encourages taking the set $\{H,\,R,\,\dot{R}\}$ as the fundamental set of dynamical variables. If one sets $\kappa=1$, then all quantities are dimensionless.

Consider a spatially flat FLRW universe devoid of matter. In this case the Friedmann equation (\ref{fe1}) and the Raychaudhuri equation (\ref{fe2}) are not independent; the latter is obtained from the former by a time derivative. Define the dynamical quantities
\begin{equation}
    x\equiv H,\quad y\equiv R,\quad z\equiv\dot{R}\;.
\end{equation}
The Friedmann equation (\ref{fe1}) provides a constraint between the dynamical variables which can be used to eliminate $z$
\begin{equation}\label{const_0}
    z=\frac{yf'(y)-f(y)-6f'(y)x^2}{6xf''(y)}\;,
\end{equation}
leaving $\{x,\,y\}$ as the only two independent dynamical variables consisting a 2-dimensional phase space. Using the Friedmann equation (\ref{fe2}) again and the definition of the Ricci scalar one can write the dynamical system as
\begin{eqnarray}
&& \dot{x}=-2x^2 +\frac{y}{6}\;,\label{dsa_01}\\
&& \dot{y}=\frac{(y-6x^2)f'(y)-f(y)}{6xf''(y)}\;,\label{dsa_02}
\end{eqnarray}
It is clear from the above that this formulation should not be used in the limit $f''\rightarrow0$ (GR limit). The entire 2-dimensional plane $x$-$y$ is not physical. The requirement $x^2>0$ singles out the physical region of the phase space as
\begin{equation}
    6xf''(y)z(x,y)-yf'(y)-f(y)<0.
\end{equation}

A complete analysis of the phase space, in particular the knowledge about the physically viable region of the phase plane,  requires the knowledge of the function $f$. Nevertheless the dynamical system comprised of Eqs.(\ref{dsa_01},\ref{dsa_02}) was used in Ref.\cite{deSouza:2007zpn} in an attempt to visualise the generic phase space features of $f(R)$ theories. The fixed points of this dynamical system, given by the condition $\{\dot{x},\dot{y}\}=\{0,0\}$ are either de-Sitter or Minkowski solutions. From Eqs.(\ref{dsa_01},\ref{dsa_02}) the fixed points are given by
\begin{equation}\label{dsfp}
    x_* =\pm\sqrt{\frac{y_*}{12}},\quad y_* f'(y_*)-2f(y_*)=0\;.
\end{equation}

$y_*=0$ corresponds to Minkowski solutions. The second condition also arises from the trace field equation (\ref{tfe}), keeping in mind that de-Sitter vacuum solutions have a constant positive value of the Ricci curvature scalar (zero value of the Ricci scalar corresponds to the Minkowski limit). Notice that the de-Sitter solutions always come in pairs, representing an expanding and a contracting solution. 

The Jacobian of the dynamical system (\ref{dsa_01},\ref{dsa_02}) at the de-Sitter fixed points simplifies as 
\begin{equation}
    J=\left(\begin{tabular}{cc}
        $-4x_*$ & $\frac{1}{6}$ \\
        $-\frac{2f'(y_*)}{f''(y_*)}$ & $x_*$
    \end{tabular}\right).
\end{equation}
If the eigenvalues are $\lambda_{1,2}$, then
\begin{equation}
    \lambda_1 +\lambda_2 =-3x_* =\mp\frac{\sqrt{3y_*}}{2},\qquad \lambda_1 \lambda_2 =\frac{1}{3}\left(\frac{f'(y_*)}{f''(y_*)}-y_*\right)\;.
\end{equation}
The de-Sitter fixed point is stable when $\lambda_1<0,\,\lambda_2<0$. The contracting de-Sitter solution can never be stable as in this case $\lambda_1 +\lambda_2 >0$. The condition for the expanding de-Sitter solution to be a stable fixed point is
\begin{equation}\label{stab_ds}
    \frac{f'(y_*)}{f''(y_*)}>y_* \Leftrightarrow \frac{f'(R_*)}{f''(R_*)}>R_*\;.
\end{equation}

\begin{figure}[h]
\begin{minipage}{12pc}
\includegraphics[scale=0.4]{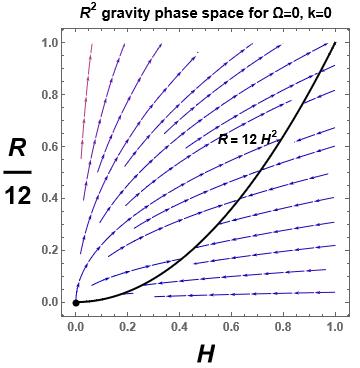}
\caption{\label{fig:1}Phase space of spatially flat vacuum solutions of $R^2$ gravity. Starobinsky's inflationary solution is represented by the whole attracting curve $R=12H^2$.}
\end{minipage}
\hspace{2pc}
\begin{minipage}{12pc}
\includegraphics[scale=0.4]{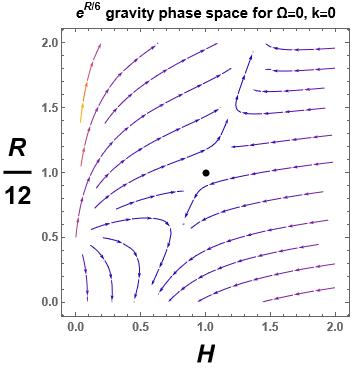}
\caption{\label{fig:2}Phase space of spatially flat vacuum solutions of $e^{R/6}$ gravity. The black dot at the center is a saddle fixed point that correspond to an inflationary solution.}
\end{minipage} 
\end{figure}

Let us consider two simple examples.
\begin{itemize}
    \item $f(R)\propto R^n$ ($n>1$): The fixed point condition (\ref{dsfp}) gives 
    \begin{equation}
        (n-2)R^n =0\;,
    \end{equation}
which is satisfied either by $R=0$ (Minkowski solution) for arbitrary $n$ or by $n=2$ for arbitrary $R$ (which encompasses both de-Sitter and Minkowski solutions). For $n=2$ ($R^2$ gravity) it is straightforward to check that $\lambda_1 \lambda_2=0$ i.e., one of the Jacobian eigenvalues must be zero. This is because in this case the De-Sitter fixed points belong to a one-parameter family of solutions. In fact, the second condition in Eq.\ref{dsfp} is identically satisfied, which means for $R^2$ gravity the parabola $y=12x^2$ in the $x$-$y$ plane constitutes a line of fixed points. In other words $R^2$ gravity has a De-Sitter point for all values of $R$. The other eigenvalue must be negative to satisfy $\lambda_1 +\lambda_2<0$. Therefore the whole curve $y=12x^2$ must be attractive in nature. The phase space is shown in Fig.\ref{fig:1}
    
\item $f(R)=e^{\alpha R}$ ($\alpha>0$): The fixed point condition (\ref{dsfp}) shows that there exists a De-Sitter fixed point at $(H,R)=\left(\frac{1}{\sqrt{6\alpha}},\frac{2}{\alpha}\right)$ (only the expanding solution is considered). However, putting this value at the stability condition (\ref{stab_ds}) shows that this fixed point is not a stable fixed point but instead a saddle. There is no Minkowski solution. The phase space is shown in Fig.\ref{fig:2}
\end{itemize}

The very basic dynamical system formulation for $f(R)$ gravity presented in this section has traditionally been employed in the context of inflation in the early universe \cite{Starobinsky:1980te,Capozziello:1993xn}, which is mostly modelled as an exponential expansion of a matter-free universe. As we have seen, this formulation proves to be particularly well-suited for this use, as the phase space is a 2-dimensional plane with the actual physically relevant quantities as the coordinates. This advantage will disappear once we consider the global spatial curvature and/or matter (i.e., terms which has an explicit scale factor dependence). Later on, as it became clear that $f(R)$ gravity theories also have immense use in the late universe context as an alternative to dark energy, ignoring the global spatial curvature or the matter is no longer justifiable. This is how the other dynamical system formulations came into the picture.
\section{The usual dynamical system formulation} \label{sec:dsa_1}
A dynamical system formulation for $f(R)$ gravity in terms of expansion-normalized (or Hubble normalized) dynamical variables was first introduced in Ref.\cite{Carloni:2004kp}, based on a paper by Ellis and Goliath \cite{Goliath:1998na}. In this formulation the dynamical variables are
\begin{equation}\label{dynvars_1}
    x=\frac{\dot{F}}{HF},\quad y=\frac{R}{6H^2},\quad z=\frac{f}{6FH^2},\quad \Omega=\frac{\kappa\rho}{3FH^2},\quad K=\frac{k}{a^2 H^2},
\end{equation}
The dynamical variables are simply obtained from the Friedmann equation (\ref{fe1}) by dividing both sides by $3H^2$ and are constrained by the simple algebraic equation 
\begin{equation}\label{const_1}
    -x+y-z-K+\Omega=1
\end{equation}
Choosing to eliminate $K$ using the Friedmann constraint, the dynamical system can be expressed as
\begin{eqnarray}
&& \frac{dx}{d\tau} = -4z -2x^2 - (z+2)x +2y +\Omega(x +1 -3w),\label{dsa_11}
\\
&& \frac{dy}{d\tau} = y[2\Omega -2(z-1) + x(\Gamma-2)],\label{dsa_12}
\\
&& \frac{dz}{d\tau} = z(-2z +2\Omega -3x +2) + xy\Gamma,\label{dsa_13}
\\
&& \frac{d\Omega}{d\tau} = \Omega(2\Omega -3x -2z -3w -1),\label{dsa_14}
\end{eqnarray}
where $\tau=\ln a$ and the auxiliary quantity $\Gamma=\Gamma(R)$ is defined as
\begin{equation}
    \Gamma(R) \equiv \frac{d\ln R}{d\ln F} = \frac{F}{RF'}\,.
    \label{Gamma}
\end{equation}
Given a functional form for $f(R)$, one can invert the relation
\begin{equation}\label{yz_f}
    \frac{y}{z}=\frac{RF}{f}
\end{equation}
(provided of course that it is invertible) to determine $R=R(y/z)$ and correspondingly find $\Gamma=\Gamma(y/z)$, so as to make the dynamical system autonomous.

This formulation has some clear benefits as compared to the one presented in Sec.\ref{sec:dsa_0}:
\begin{itemize}
    \item In the formulation of Sec.\ref{sec:dsa_0} one required to adopt a natural unit system ($\kappa=1$) to make the dynamical variables dimensionless. In the present formulation the dynamical variables are by construction dimensionless.
    \item The constraint equation (\ref{const_1}) has a much simpler form than Eq.(\ref{const_0}).
    \item In the formulation of Sec.\ref{sec:dsa_0} the fixed points was necessarily De-Sitter solutions (or Minkowski solutions in special cases). In the present formulation even scaling solutions $a\sim t^n$ may arise as fixed points (e.g. for the spatially flat case this corresponds to $y=2-\frac{1}{n}$).
\end{itemize}

\begin{figure}[h]
\includegraphics[scale=0.6]{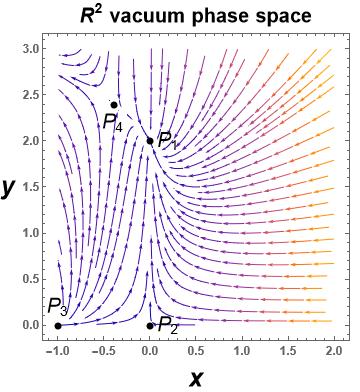}
\hspace{2pc}
\begin{minipage}[b]{14pc}\caption{\label{fig:3}Phase space of vacuum solutions of $R^2$ gravity. Starobinsky's inflationary solution is represented by the stable fixed point $P_1$.}
\end{minipage}
\end{figure}

All the phase trajectories corresponding to a possible vacuum cosmology reside on the invariant submanifold $\Omega=0$ (An invariant submanifold is one such that no phase trajectory can cross it and there are phase trajectories residing entirely on it. An invariant submanifold divides the entire phase space in two disjoint sectors). It is interesting to consider the phase space of vacuum solutions of $R^n$ theories in the present formulation to illuminate what is the difference of this formulation with the earlier one of Sec.\ref{sec:dsa_0}. Firstly, it is to be noted that such minimal theories have the added advantage of introducing a new constraint \cite{Carloni:2004kp}
\begin{equation}\label{const_mon}
    y-nz=0.
\end{equation}
Also, for $R^n$ theories the auxiliary quantity $\Gamma$ is constant;
\begin{equation}
    \Gamma=\frac{1}{n-1}
    \label{Gamma1}
\end{equation}
Therefore the phase space of vaccum $R^n$ cosmology can be reduced to the 2-dimensional $x$-$y$ (or $x$-$z$) plane irrespective of the global spatial curvature. Notice the advantage as compared to the formulation of Sec.\ref{sec:dsa_0}. Had we not considered the spatially flat case in Sec.\ref{sec:dsa_0} the phase space of vacuum solutions would have been 3-dimensional, unlike the present formulation where the phase space is always 2-dimensional irrespective of the global spatial curvature. In particular, the phase space of vacuum solutions of $R^2$ gravity is given by the dynamical system
\begin{eqnarray}
&& \frac{dx}{d\tau} = -2x-\frac{1}{2}y-2x^2\;,
\\
&& \frac{dy}{d\tau} = -x+2y-y^2\;.
\end{eqnarray}
There are four fixed points of the above system, listed in table \ref{tab:1}. The corresponding phase space is shown in Fig.\ref{fig:3}.
\begin{table}[h]
    \centering
    \begin{tabular}{|c|c|c|c|}
        \hline
        Fixed point & $(x,y,K)$ & Stability & Cosmology \\
        \hline
        $P_1$ & $(0,2,0)$ & Stable & $H=const.$ \\
        \hline
        $P_2$ & $(0,0,-1)$ & Saddle & $a\sim t^{1/2}$ \\
        \hline
        $P_3$ & $(-1,0,0)$ & Unstable & $a\sim t^{1/2}$ \\
        \hline
        $P_4$ & $\left(-\frac{2}{5},\frac{12}{5},\frac{3}{5}\right)$ & Saddle & $a\sim\frac{1}{(t_s -t)^{5/3}}$, $t_s=\frac{5}{2H(t=0)}$\\
        \hline
    \end{tabular}
    \caption{Fixed points in the phase space of vacuum cosmology in $R^2$ gravity.}
    \label{tab:1}
\end{table}
Note that the scaling cosmology represented by $P_3$, which is a fixed point for the spatially flat case, was not obtained by the formulation of Sec.\ref{sec:dsa_0}. This clearly shows the advantage of the present formulation for a detailed analysis of the phase space. Moreover, Starobinski's inflationary solution, which corresponded to a whole curve in the formulation of Sec.\ref{sec:dsa_0}, is represented in this formulation by the single point $P_1$. This helps to get a clearer understanding of the stability nature of this solution.
Apart from $R^n$ gravity, however there are no other $f(R)$ theory that gives an additional constraint like Eq.(\ref{const_mon}), and consequently we would not be able to reduce the dimensionality of the phase space like we could do for $R^n$ theory.

The limitation of this formulation is that it's success is crucially dependent on the invertibility of the relation (\ref{yz_f}) to find $R=R(y/z)$. This renders the present formulation useless in many situations. A particularly important example of this is the Hu-Sawicki $f(R)$ model that was purposefully built as an $f(R)$ alternative to dark energy \cite{Hu:2007nk}. The generic form of the model is given by a Lagrangian of the form 
\begin{equation}\label{HS}
    f(R)=R-m^2 \frac{C_1 \left(R/m^2\right)^n}{C_2 \left(R/m^2\right)^n +1}\;,
\end{equation}
where $C_1,\,C_2,\,n$ are free parameters and $m^2$ is a characteristic mass scale which can be taken, for example, to be $m^2 = \kappa\rho_{m0}/3 = \Omega_{m0} H_0^2$. For this model the relation (\ref{yz_f}) becomes
\begin{equation}
\frac{y}{z}=\frac{Rf'}{f}=\frac{(1+C_2 r^n)^2 - nC_1 r^{n-1}}{(1+C_2 r^n)(1 - C_1 r^{n-1} + C_2 r^n)}\;,
\end{equation}
where $r=R/m^2$. Clearly, the above expression is in general non-invertible except for the special case $n=1$ \cite{Kandhai:2015pyr}. Therefore a generic dynamical system analysis of this model is not possible with the dynamical system formulation of this section. The alternative dynamical system formulation that is described in the next section has been proposed as a means to overcome this limitation.
\section{An alternative dynamical system formulation} \label{sec:dsa_2}
To overcome the limitation arising from the invertibility requirement, a formulation of the $f(R)$ dynamical system was presented by Carloni in Ref.\cite{Carloni:2015jla}, which is based on the following variables
\begin{equation}\label{dynvars_2}
   y=\frac{R}{6 H^2},\quad K=\frac{k}{a^2 H^2},\quad\Omega=\frac{\rho}{3f' H^2},\quad A=\left(\frac{H}{m}\right)^2,\quad Q=\frac{3}{2}\frac{H,_{\tau}}{H},\quad J=\frac{1}{4}\frac{H,_{\tau\tau}}{H}.
\end{equation}
In the above $m$ is some constant with a mass dimension, which is typically chosen from the parameters appearing in the particular $f(R)$ theory under consideration. Comparing with the formulation of Sec.\ref{sec:dsa_1}, we notice that the dynamical variables $y,\,K,\,\Omega$ are kept same whereas three new variables have been introduced. The function $f(R)$ enters into the definition of the dynamical variables only through the one variable, namely $\Omega$. All the other dynamical variables are completely kinematic in nature, i.e., they depend entirely on \emph{how} the universe evolves and not at all on what is it's inherent dynamics. In fact, the variables $Q,\,J$ can be related to the cosmographic deceleration parameter $q=-\frac{\ddot{a}}{aH^2}$ and jerk parameter $j=\frac{1}{aH^3}\frac{d^3 a}{dt^3}$ as
\begin{eqnarray}
&& Q=-\frac{3}{2}(1+q)\,, \label{Q}\\
&& J=\frac{1}{4}(1+q-q^2 +j)\;. \label{J}
\end{eqnarray}
The Friedmann equation and the definition of the Ricci scalar provides the following constraints
\begin{eqnarray}
&& \Omega-K+y-\mathbf{X}-\left[\left(1+\frac{Q}{9}\right)Q-\frac{y}{2}+J+1\right]\mathbf{Y}=1, \label{ricci_const_1}\\
&& y-K-\frac{2}{3}Q=2. \label{fried_const_1}
\end{eqnarray}
where two auxiliary quantities have been introduced
\begin{equation}
    \mathbf{X}=\frac{f}{6H^2 f'},\quad\mathbf{Y}=\frac{24H^2 f''}{f'}\;.
\end{equation}
Using the constraints to eliminate the cosmographic variables $Q,\,J$, the dynamical system can be expressed as
\begin{eqnarray}
&& \frac{dy}{d\tau}=2y(K-y+2)-\frac{4}{\mathbf{Y}}(\mathbf{X}+K-y-\Omega+1), \label{dsa_21}\\
&& \frac{d\Omega}{d\tau}=\Omega(2-3w+\mathbf{X}+3K-3y-\Omega), \label{dsa_22}\\
&& \frac{dK}{d\tau}=2K(K-y+1), \label{dsa_23}\\
&& \frac{dA}{d\tau}=-2A(2+K-y). \label{dsa_24}
\end{eqnarray}
Comparing with the dynamical system formulation of Sec.\ref{sec:dsa_1}, we see that both approaches produce a 4-dimensional phase space as expected. However, unlike the earlier formulation that has one auxiliary quantity $\Gamma$ explicitly depending on the functional form of $f(R)$, this alternative formulation has two such auxiliary quantities, namely $\mathbf{X}$ and $\mathbf{Y}$. The possibility of obtaining a closed autonomous dynamical system depends on the ability to express these auxiliary quantities in terms of the dynamical variables. In the usual formulation this crucially depends on the invertibility of the relation $\frac{y}{z}=\frac{Rf'}{f}$ to find $R=R(y/z)$. However, noting that
\begin{equation}
    H^2 = m^2 A,\quad R=6H^2 y =6m^2 Ay
\end{equation}
and that $\mathbf{X},\,\mathbf{Y}$ are functions of $H^2$ and $R$ only, there is no such limitation in this alternative formulation. Depending on $f(R)$, the forms of the functions $\mathbf{X}(A,y)$ and $\mathbf{Y}(A,y)$ may come out as quite complicated. Nevertheless, this alternative formulation can be used in principle to consider any form of $f(R)$ whatsoever. Here lies the advantage of this formulation. For example, let us consider again the generic form of the Hu-Sawicki model (Eq.(\ref{HS})) and take $m^2$ in the definition of $A$ to be the same as the one appearing in the model. Straightforward calculation gives
\begin{eqnarray}
&& \mathbf{X} = y \frac{(1 + C_2(6Ay)^n)(1 - C_1(6Ay)^{n-1} + C_2(6Ay)^n)}{(1 + C_2(6Ay)^n)^2 - nC_1(6Ay)^{n-1}},\\
&& \mathbf{Y} = \frac{4}{y} \frac{nC_1(6Ay)^{n-1}[(n+1)C_2(6Ay)^n - (n-1)]}{(1+C_2(6Ay)^n)[(1+(6Ay)^n)^2 - nC_1(6Ay)^{n-1}]}.
\end{eqnarray}
The dynamical equations becomes very complicated in form. Their explicit expressions can be found in Ref.\cite{Carloni:2015jla}, so we do not find it necessary to write them here. Nevertheless, the present formulation in principle allows for a dynamical system analysis of Hu-Sawicki model in it's generic form, unlike the formulation in the previous section, which allows for only the special case of $n=1$. 

The formulation presented in this section, as well as the ones given in Sec.\ref{sec:dsa_0} and Sec.\ref{sec:dsa_1} works for a top-down approach, i.e. a functional form of $f(R)$ must be specified a-priori. Once an $f(R)$ is given, it's solution dynamics can be understood by carrying out a dynamical analysis by one of the above formulations. In a bottom-up approach the the functional form of $f(R)$ is not known a-priori, but the resulting cosmology is required to satisfy some condition. The approach presented in the next section has grown out of an attempt to utilize the dynamical system technique in the bottom-up approach.
\section{The form-independent dynamical system approach} \label{sec:dsa_3}
In the bottom-up approach of doing cosmology one first specifies a cosmology ($a(t)$) and, given a matter content, tries to find out what functional form of $f(R)$ gravity can produces the specified dynamics. This can be achieved via various reconstruction methods \cite{Nojiri:2009kx,Dunsby:2010wg,He:2012rf,Lee:2017lud,Sami:2017nhw,Qiu:2012np,Chiba:2018cmn}. Then one can carry out a dynamical systems analysis of the reconstructed $f(R)$ model. The problem is that more often than not the $f(R)$ cannot be reconstructed as a compact functional form but only as a series or in terms of special functions. It would be great if we could somehow avoid the entire reconstruction exercise but still get a glimpse of the generic dynamical features of whatever the underlying $f(R)$ theory is. This is what is meant by the term \emph{form-independence}, i.e,  one need not specify a functional form of $f(R)$ a-priori. The essential idea is to somehow project the cosmographic requirement used to reconstruct the $f(R)$ gravity as an algebraic constraint over a suitably defined phase space. Below we outline two such approaches and illustrate them with the example of all possible $f(R)$ dynamics that can exactly mimic the $\Lambda$CDM expansion history. 
\subsection{Approach I}
One such approach, built up on the dynamical system formulation presented in Sec.\ref{sec:dsa_1}, was recently introduced by Chakraborty et al \cite{Chakraborty:2021jku}. The formulation employs the same dynamical variables as in Eq.(\ref{dynvars_1}). Additionally it also makes use of three cosmographic parameters, namely the deceleration, jerk and snap parameters \cite{Dunajski:2008tg}:
\begin{equation}
    q \equiv -\frac{1}{aH^2}\frac{d^2 a}{dt^2}, \qquad j \equiv \frac{1}{aH^3}\frac{d^3 a}{dt^2}, \qquad s \equiv \frac{1}{aH^4}\frac{d^4 a}{dt^2},
\end{equation}
which are related to each other by
\begin{eqnarray}
&& j = 2q^2 + q - \frac{dq}{d\tau}\,,
\\
&&  s = \frac{dj}{d\tau} - j(2 + 3q)\,.
\end{eqnarray}
The main idea of the formulation is that, since the cosmographic parameters are also dimensionless, they can also be used as dynamical variables. Let us consider an extended phase space spanned by the earlier dynamical variables plus the cosmographic parameters. The Friedmann equation and the definition of the Ricci scalar provides two constraints that can be used to reduce the phase space dimensionality by two.. We note that in the usual formulation the term $\Gamma$, which explicitly carries information about the function $f(R)$, appears in the dynamical equations of $y$ and $z$. Therefore we choose to eliminate $y$ and $z$ using the two constraints. The definition of Ricci scalar and the Friedmann equation can be expressed as
\begin{eqnarray}
&& y=1-q+K,\label{ricci_const}\\
&&  z=-x+\Omega-q\label{fried_const}.
\end{eqnarray}
The resulting dynamical system for a barotropic perfect fluid of equation of state $w$ is
\begin{eqnarray}
&& \frac{dx}{d\tau} = -x(x -q) +2(x +K +q) -3\Omega(1 +w) +2\;, \label{dsa_31}
\\
&& \frac{d\Omega}{d\tau} = -\Omega(x -2q +1 +3w)\;, \label{dsa_32}
\\
&& \frac{dK}{d\tau} = 2qK\;, \label{dsa_33}
\\
&& \frac{dq}{d\tau} = 2q^2 +q -j\;, \label{dsa_34}
\\
&& \frac{dj}{d\tau} = j(2 +3q) +s\;. \label{dsa_35}
\end{eqnarray}
Since $f(R)$ gravity is a fourth order theory of gravity, i.e,. the field equations contain terms including up to fourth derivative of the metric, cosmographic parameters of order higher than $s$ cannot appear in the field equations. The above dynamical system has a much simpler form as compared to the dynamical system of Sec.\ref{sec:dsa_1} and also does not require to explicitly specify the functional form of the underlying $f(R)$ gravity to make the system autonomous. The price to pay is that one has to now specify a relation between the cosmographic parameters $q,\,j,\,s$, which is essentially the same as specifying a cosmology $a(t)$ in the reconstruction methods. 

As mentioned in the introduction, physical viability of any $f(R)$ gravity requires $F>0$ throughout the physically relevant region of the phase space and $F'\geq 0$ ($F'=0$ corresponding to the special case $f(R)=R+\Lambda$) at least in the neighbourhood of the fixed point corresponding to the matter dominated epoch. It can be noted that
\begin{equation}
    xy\Gamma = -2(1+y) +6(x+z+\Omega) + j + 3q - 2\,;.
\end{equation}
Eliminating $y$ and $z$ on the right hand side by using the constraints (\ref{ricci_const}) and (\ref{fried_const}) and using the definition of $\Gamma$ one can write
\begin{equation}
    \frac{1}{y\Gamma} = \frac{6F'H^2}{F} = \frac{x}{12\Omega -2K -q +j -6}\,.
\end{equation}
Assuming the condition $F>0$ is met, demanding $F'\geq 0$ puts the following constraint on the phase phase:
\begin{equation}\label{phys_const}
    \frac{x}{12\Omega -2K -q +j -6} \geq 0\;.
\end{equation}
The submanifold $x=0$ corresponds to the GR limit ($F'=0$). 

As an example let us consider the possibility of reconstructing $f(R)$ so as to reproduce exactly the same background evolution as produced by the observationally successful $\Lambda$CDM model. This reconstruction exercise is in principle possible, but the reconstructed $f(R)$ can be written only in terms of Hypergeometric functions \cite{Dunsby:2010wg,He:2012rf}. $\Lambda$CDM cosmology can be specified by the simple cosmographic requirement \cite{Dunajski:2008tg,Capozziello:2008qc}
\begin{equation}\label{LCDM}
   j=K+1\,.
\end{equation}
This represents a 4-dimensional submanifold in the 5-dimensional phase space given by Eqs.(\ref{dsa_31},\ref{dsa_32},\ref{dsa_33},\ref{dsa_34},\ref{dsa_35}). The dynamical system for all possible $f(R)$ theories that can exactly mimic the $\Lambda$CDM cosmology can be written as
\begin{eqnarray}
&& \frac{dx}{d\tau} = -x(x -q) +2(x +K +q) -3\Omega +2\;, \label{dsa_lcdm_11}
\\
&& \frac{d\Omega}{d\tau} = -\Omega(x -2q +1)\;, \label{dsa_lcdm_12}
\\
&& \frac{dK}{d\tau} = 2qK, \label{dsa_lcdm_13}
\\
&& \frac{dq}{d\tau} = 2q^2 +q -K -1\;. \label{dsa_lcdm_14}
\end{eqnarray}
Fixed points of the above system, along with their nature of linear stability, are listed in table \ref{tab:2}.
\begin{table}[!h]
    \centering
    \begin{tabular}{|c|c|c|c|}
        \hline 
        Fixed & Coordinates & Stability & Cosmological solution \\
        Point & $(x^*,\Omega^*,K^*,q^*)$ & Nature & \\
        \hline
        $P_1$ & $(0,0,0,-1)$ & Saddle & Scalaron dominated De-Sitter ($H=$const.) \\
        \hline
        $P_2$ & $(1,0,0,-1)$ & Attractor & Scalaron dominated De-Sitter ($H=$const.) \\
        \hline
        $P_3$ & $(0,1,0,\frac{1}{2})$ & Saddle & Matter dominated power law ($a\sim t^{2/3}$) \\
        \hline
        $P_4$ & $(\frac{5-\sqrt{73}}{4},0,0,\frac{1}{2})$ & Repeller & scalaron dominated power law ($a\sim t^{2/3}$) \\
        \hline
        $P_5$ & $(\frac{5+\sqrt{73}}{4},0,0,\frac{1}{2})$ & Saddle & Scalaron dominated power law ($a~t^{2/3}$) \\
        \hline
        $P_6$ & $(0,0,-1,0)$ & Saddle & Milne solution ($a\sim t$) \\
        \hline
        $P_7$ & $(2,0,-1,0)$ & Saddle & Milne solution ($a\sim t$) \\
        \hline
    \end{tabular}
    \caption{Fixed points in the phase space of $f(R)$ cosmologies that are same as $\Lambda$CDM.}
    \label{tab:2}
\end{table}
The phase trajectories for all such spatially flat cosmologies lie on the invariant submanifold $K=0$, where the phase space can be reduced as
\begin{eqnarray}
&& \frac{dx}{d\tau} = -x(x -q) +2(x +q) -3\Omega +2\;, \label{dsa_lcdm_flat_11}
\\
&& \frac{d\Omega}{d\tau} = -\Omega(x -2q +1)\;, \label{dsa_lcdm_flat_12}
\\
&& \frac{dq}{d\tau} = 2q^2 +q -1\;. \label{dsa_lcdm_flat_13}
\end{eqnarray}
We notice that for the spatially flat case the $q$-equation decouples which leads to two new invariant submanifolds: a submanifold $q=-1$ consisting of accelerated cosmological solutions and a submanifold $q=\frac{1}{2}$ consisting of decelerated cosmological solutions. The fixed points $P_1$ and $P_2$ reside on the $q=-1$ submanifold and the fixed points $P_3$, $P_4$ and $P_5$ sit on the $q=\frac{1}{2}$ submanifold. Linear stability analysis reveals that the deceleration submanifold $q=\frac{1}{2}$ is a repelling one while the acceleration submanifold $q=-1$ is an attracting one, which is consistent with the fact that the universe has transitioned from a decelerating epoch to an accelerating epoch. Since the deceleration parameter $q$ is monotonically decreasting with time between $q=\frac{1}{2}$ and $q=-1$, it is possible to represent the solutions as parametric curves $\{x(q),\Omega(q)\}$ on the 2-dimensional $x$-$\Omega$ plane (these are \emph{not} the phase trajectories, as the two equations (\ref{dsa_lcdm_flat_11}) and (\ref{dsa_lcdm_flat_12}) by themselves do not represent an autonomous system). The plot is shown in Fig.\ref{fig:4}.

It should be noted that the scalaron dominated de-Sitter future attractor $P_2$ does not satisfy the condition (\ref{phys_const}). This allows us to conclude that even if there exist possible $f(R)$ models which are able to give rise to a cosmological dynamics that is indistinguishable from the $\Lambda$CDM model at the background level, the $f(R)$-dynamics will inevitably lead to an epoch where the condition $F'>0$ is not met. However, as long as $F'>0$ for enough time around the matter dominated fixed point $P_3$, there should not be any theoretical pathology in the model.
\subsection{Approach II}
A different approach can be built up with the dynamical system formulation presented in Sec.\ref{sec:dsa_2}. We utilise the same dynamical variables as defined in Eq.(\ref{dynvars_2}). Unlike in the previous subsection, in this approach we do not need to extend the phase space by incorporating cosmographic parameters by hand, as the cosmographic parameters are, in some way, already incorporated in the definitions of the dynamical variables (see Eqs.(\ref{Q},\ref{J})). As mentioned in Eq.(\ref{LCDM}), the $\Lambda$CDM expansion history is specified by the cosmographic condition
\begin{equation}
    j = K + 1\;.
\end{equation}
This represents a 3-dimensional submanifold in the 4-dimensional phase space given by Eqs.(\ref{dsa_21},\ref{dsa_22},\ref{dsa_23},\ref{dsa_24}). To find the equation of this surface we first invert the relations in Eqs.(\ref{Q},\ref{J}) to find
\begin{equation}
    q = -1 - \frac{2}{3}Q, \quad j = 1 + 2Q + \frac{4}{9}Q^2 + 4J\;.
\end{equation}
The cosmographic condition (\ref{LCDM}) can now be written as
\begin{equation}\label{LCDM_1}
    J = \frac{1}{4}\left(K - 2Q - \frac{4}{9}Q^2\right).
\end{equation}
Using this cosmographic constraint along with the Ricci constraint (\ref{ricci_const_1}), the Friedmann constraint (\ref{fried_const_1}) can be expressed as 
\begin{equation}\label{LCDM_2}
\frac{4}{\mathbf{Y}}(\mathbf{X} + K - y - \Omega + 1) + y - 2K - 2 = 0. 
\end{equation}
which is the equation for the 3-dimensional submanifold containing phase trajectories that correspond to $\Lambda$CDM-mimicking $f(R)$ cosmologies, i.e., cosmologies that are indistinguishable from the $\Lambda$CDM model at the background level. Using Eq.(\ref{LCDM_2}) to eliminate $\Omega$, we can find the dynamical system governing the phase dynamics on this submanifold.
\begin{eqnarray}
&& \frac{dy}{d\tau} = -(2y-1)(y-K-2)-K\;, \label{dsa_lcdm_21}
\\
&& \frac{dK}{d\tau} = -2K(y-K-1)\;, \label{dsa_lcdm_22}
\\
&& \frac{dA}{d\tau} = 2A(y-K-2)\;. \label{dsa_lcdm_23}
\end{eqnarray}
Fixed points of the above system, along with their nature of linear stability, are listed in table \ref{tab:3}.
\begin{table}[!h]
    \centering
    \begin{tabular}{|c|c|c|c|}
        \hline 
        Fixed & Coordinates & Stability & Cosmological solution \\
        Point & $(y^*,K^*,A^*)$ & Nature & \\
        \hline
        $\mathcal{P}$ & $\left(\frac{1}{2},0,0\right)$ & Saddle & Minkowski ($H=0$) \\
        \hline
        $\mathcal{L}$ & $(2,0,A)$ & Attractor & De-Sitter ($H=$const.) \\
        \hline
        $\mathcal{Q}$ & $(0,-1,0)$ & Saddle & Milne solution ($a\sim t$) \\
        \hline
    \end{tabular}
    \caption{Fixed points in the phase space of $f(R)$ cosmologies that are same as $\Lambda$CDM. $\mathcal{L}$ is actually a line of fixed points.}
    \label{tab:3}
\end{table}

Spatially flat cosmologies reside on the invariant submanifold $K=0$, on which the dynamical system can be further reduced as
\begin{eqnarray}
&& \frac{dy}{d\tau} = -(2y-1)(y-2), \label{dsa_lcdm_flat_21}
\\
&& \frac{dA}{d\tau} = 2A(y-2).\label{dsa_lcdm__flat_22} 
\end{eqnarray}
The corresponding phase portrait is shown in Fig.\ref{fig:5}. We notice that for the spatially flat case the $y$-equation decouples which leads to two new invariant submanifolds: a submanifold $y=2$ consisting of de-Sitter cosmological solutions and a submanifold $y=\frac{1}{2}$ consisting of power law cosmological solutions $a\sim t^{2/3}$. Linear stability analysis reveals that the deceleration submanifold $y=\frac{1}{2}$ is a repelling one while the acceleration submanifold $y=2$ is an attracting one. In fact, as is also clear from Fig.\ref{fig:5} the submanifold $y=2$ is actually line of De-Sitter fixed points $\mathcal{L}$.
\subsection{Comparison between the two approaches}
\begin{figure}[h]
\begin{minipage}{12pc}
\includegraphics[scale=0.4]{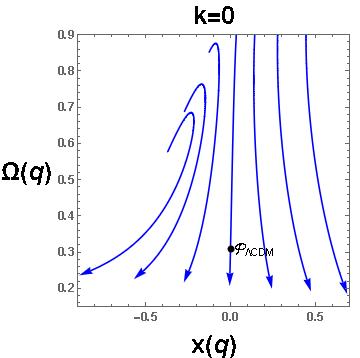}
\caption{\label{fig:4} Some typical spatially flat $f(R)$ cosmologies mimicking the $\Lambda$CDM evolution history are found by approach I are shown as parametric curves $\{x(q),\Omega(q)\}$ on the $x$-$\Omega$ plane. The deceleration parameter $q$ is monotonically decreasing with time between the invariant submanifolds $q=\frac{1}{2}$ and $q=-1$, as is clear from Eq.(\ref{dsa_lcdm_flat_13}. Solution dynamics is represented by the arrows on the curves. The point $\mathcal{P}_{\Lambda\rm{CDM}}$ corresponds the present day epoch in $\Lambda$CDM model.}
\end{minipage}
\hspace{2pc}
\begin{minipage}{12pc}
\includegraphics[scale=0.4]{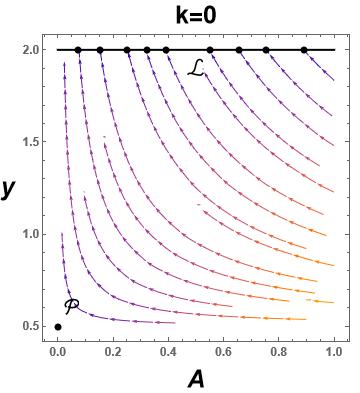}
\caption{\label{fig:5} Phase portrait of spatially flat $\Lambda$CDM mimicking $f(R)$ cosmologies found by approach II. The saddle fixed point $\mathcal{P}$ corresponds to a Minkowski solution and the line $\mathcal{L}$ represent a line of De-Sitter attractors. The figure clearly shows transitions from decelerated cosmology ($y=\frac{1}{2}$) to accelerated cosmology ($y=2$).}
\end{minipage} 
\end{figure}

An apparent advantage of approach II over the approach I is that it reduces the phase space dimensionality by one. We note that the dynamical system given by Eqs.(\ref{dsa_lcdm_11},\ref{dsa_lcdm_12},\ref{dsa_lcdm_13},\ref{dsa_lcdm_14}) and the dynamical system given by Eqs.(\ref{dsa_lcdm_21},\ref{dsa_lcdm_22},\ref{dsa_lcdm_23}) both correspond to $\Lambda$CDM mimicking $f(R)$ cosmologies. However, in the former case the phase space is 4-dimensional, whereas in the later it is 3-dimensional. In particular for the spatially flat $\Lambda$CDM mimicking $f(R)$ cosmologies, the phase space is 3-dimensional in approach I, whereas in approach II it is 2-dimensional, allowing us to plot a phase portrait (Fig.\ref{fig:5}). However, this advantage is only \emph{apparent} because approach II is actually \emph{not} form-independent. If we want to know the complete fixed point structure, we need to calculate the $(\Omega,Q,J)$ coordinates of the fixed points using the constraint equations (\ref{LCDM_2}), (\ref{fried_const_1}), (\ref{ricci_const_1}). Although the autonomous dynamical system given by Eqs.(\ref{dsa_lcdm_21},\ref{dsa_lcdm_21},\ref{dsa_lcdm_21}) is free of any explicit reference to the functional form of $f(R)$, both the constraints (\ref{LCDM_2}) and (\ref{fried_const_1}) carry explicit dependence on the functional form of $f(R)$ entering through the quantities $\mathbf{X}$ and $\mathbf{Y}$. The complete coordinate of a fixed point cannot be calculated without first knowing the functional form of $f(R)$.

Secondly, we note that in approach II, the phase space coordinates $(y,K,A)$ are all completely kinematic in nature. This is in contrast to approach I where the coordinate $x$ explicitly characterize the deviation from GR ($x\rightarrow0$ in GR limit). Therefore Fig.\ref{fig:5} does not exactly help us understand the difference between the original $\Lambda$CDM cosmology and the $\Lambda$CDM mimicking $f(R)$ cosmologies. For example, the approach of the previous subsection reveals that there are three distinct fixed points with different values of the $x$-coordinate on the deceleration submanifold and two distinct fixed points with different values of the $x$-coordinates on the acceleration submanifold. Information about this degeneracy is completely absent in approach II.  

Apart from allowing us for a form-independent dynamical analysis, another important merit possessed by both these approaches is that they are regular at the GR limit ($F'\rightarrow0$), whereas the actual formulations in Sec.\ref{sec:dsa_1} and Sec.\ref{sec:dsa_3} were singular at the GR limit. 

\section{Compact Dynamical Systems Formulation}
\label{sec:compact}

An alternative dynamical systems formulation for $f(R)$ gravity was introduced by \cite{Abdelwahab:2011dk} to address the issues present in the standard expansion-normalized formulation. Since the dynamical variables in Sec. \ref{sec:dsa_2} are defined such that $H\neq 0$, finite cosmological evolutions involving static states, cosmological bounces or recollapses are not possible in this formulation. In order to pull any solutions at infinity into a finite phase space, a new positive normalisation is defined to confine all global dynamics to a simply-visualised region.

In the compact dynamical systems formulation, the Friedmann equation (\ref{fe1}) is rewritten as 
\begin{equation}
\left(3 H+\frac{3}{2} \frac{\dot{F}}{F}\right)^{2}+\frac{3 }{2} \left(\frac{f}{F}+\frac{6k}{a^2}\right) = \frac{3\rho_m}{F} + \frac{3}{2}R + \left(\frac{3}{2}\frac{\dot{F}}{F}\right)^2,
\end{equation}
allowing for the introduction of the normalisation variable $D$ defined as 
\begin{equation}
D^{2}=\left(3 H+\frac{3}{2} \frac{\dot{F}}{F}\right)^{2}+\frac{3 }{2} \left(\frac{f}{F}+\frac{6k}{a^2}\right).
\end{equation}
From this we can define a set of new dynamical variables
\begin{equation}
\begin{aligned}
& \bar{x}=\frac{3}{2} \frac{\dot{F}}{F} \frac{1}{D}, \quad \bar{y}=\frac{3}{2} \frac{R}{D^{2}}, \quad \bar{z}=\frac{3}{2} \frac{f}{F} \frac{1}{D^{2}}, \\
& \bar{\Omega}=\frac{3 \rho_{m}}{F} \frac{1}{D^{2}}, 
\quad \bar{Q}=\frac{3 H}{D}, \quad \bar{K} = \frac{9k}{a^2}\frac{1}{D^2}.
\label{var_def}
\end{aligned}
\end{equation}

We now get two independent constraint equations, the first coming from the reformulated Friedmann equation, 
\begin{equation}
    1 = \bar{\Omega} + \bar{y} + \bar{x}^2,
    \label{friedconstraint}
\end{equation}
and the second from the definition of $D$ 
\begin{equation}
    1 = (\bar{Q}+\bar{x})^2 + \bar{z} + \bar{K}.
    \label{compactconstraint}
\end{equation}
If we restrict ourselves to the case of nonnegative spacetime curvature ($R\geq0$) and nonnegative global curvature for the spatial section ($k\geq0$), then, along with the physical viability condition $F>0$, the constraint equations above define a compact phase space bound by
\begin{equation}
\begin{aligned}
-1 \leq \bar{x} \leq 1, \quad 0 \leq &\bar{\Omega} \leq 1, \quad-2 \leq \bar{Q} \leq 2 \\
0 \leq \bar{y} \leq 1, \quad 0 \leq &\bar{z} \leq 1, \qquad 0\leq \bar{K} \leq 1.
\end{aligned}
\end{equation}
We also introduce a normalized time variable $\bar{\tau}$ such that
\begin{equation} 
    \frac{d}{d\bar{\tau}} \equiv \frac{1}{D}\frac{d}{dt}
\end{equation}
Compact and non-compact dynamical variables and time variables are related as follows
\begin{equation}\label{var_reln}
    \begin{aligned}
    & x = 2\frac{\bar{x}}{\bar{Q}}, \quad y = \frac{\bar{y}}{\bar{Q}^2}, \quad z = \frac{\bar{z}}{\bar{Q}^2},\\
    & K = \frac{\bar{K}}{\bar{Q}^2}, \quad \Omega = \frac{\bar{\Omega}}{\bar{Q}^2},\\
    & \frac{d}{d\tau} = \frac{1}{|H|}\frac{d}{dt} = \frac{\epsilon}{Q}\frac{d}{d\bar{\tau}}.
    \end{aligned}
\end{equation}
where $\epsilon\equiv\frac{H}{|H|}$. Notice that in the last line we have written $d\tau=|H|dt$ rather than $d\tau=Hdt$ as we did earlier for the non-compact analysis. This is because for the non-compact phase space formulation the definition of the phase space time variable is different for expanding and contracting universe and earlier we have explicitly considered expanding universe scenario for which $\epsilon=1$. The definition of $\tau$ is not continuous through a bounce or a recollapse, which re-emphasizes the shortcoming of the non-compact formulation.

As in Sec. \ref{sec:dsa_1}, we can choose to eliminate two of the variables using the constraint equations and express the dynamical system as 
\begin{eqnarray}
       \frac{d\bar{y}}{d\bar{\tau}} =&& -\frac{1}{3}\bar{y}\left( (\bar{Q}+\bar{x})\Big(2\bar{y} -(1+3w)(1-\bar{x}^2-\bar{y})   +4\bar{x}\bar{Q}   \Big)-2\bar{Q} -4\bar{x}   +2 \bar{x} \Gamma(\bar{y}-1) +4\bar{x}\bar{K} \right)\,,\\
       \frac{d\bar{x}}{d\bar{\tau}}=&&\frac{1}{6}\Big(-2 \bar{x}^{2} \bar{y} \Gamma+(1-3 w)(1-\bar{x}^2-\bar{y})+2 \bar{y}+4\left(\bar{x}^{2}-1\right)\left(1-\bar{Q}^{2}-\bar{x} \bar{Q}\right)+\bar{x}(\bar{Q}+\bar{x})\nonumber\\
       &&\left((1+3 w)(1-\bar{x}^2-\bar{y})-2 \bar{y}\right)+4\bar{K}(1-\bar{x}^2)\Big)\,,\\
       \frac{d\bar{Q}}{d\bar{\tau}}= &&\frac{1}{6}\Big(-4\bar{x}\bar{Q}^3+\bar{x}\bar{Q}(5+3w)(1-\bar{x}\bar{Q})-\bar{Q}^2(1-3w)-\bar{Q}\bar{x}^3(1+3w)-3\bar{y}\bar{Q}(1+w)(\bar{Q}+\bar{x})\nonumber\\
&&+2\bar{y}(1-\Gamma \bar{Q}\bar{x}) - 2\bar{K}(1+2\bar{x}\bar{Q})\Big)\,,\\
\frac{d\bar{K}}{d\bar{\tau}}=&&-\frac{1}{3} \bar{K}\left( (\bar{Q}+\bar{x})\Big( -(1+3w)(1-\bar{x}^2-\bar{y})  +4\bar{x}\bar{Q}   +2\bar{y}\Big)+4\bar{x}(\bar{K}-1)+2 \bar{x}\bar{y} \Gamma    \right),
    \label{eq:compactdynsys}
    \end{eqnarray}
where we again make use of the auxiliary quantity $\Gamma$ (\ref{Gamma1}). We again require that $\Gamma$ be expressed in terms of the dynamical variables in order to close the system. 
This formulation benefits from the same points mentioned in Sec. \ref{sec:dsa_1}, including the dimensionless variables, fixed points leading to complex cosmological solutions and simple constraint equations allowing for the reduction in dimension of the phase space. Much like in the non-compact dynamical system, this formulation can also be extended to include the cosmographic deceleration and jerk parameters, effectively eliminating the dependence on the functional form of $f(R)$. However, this leads to a set of fairly complex propagation equations. In addition to these benefits, in the compact formalism the dynamical variables are well-defined when $H=0$, allowing for the existence of well-defined, finite trajectories and fixed points corresponding to static, expanding, collapsing and bouncing solutions. The main purpose of this formulation is the visualisation of the entire global phase space. Since all trajectories are confined to the compact region defined by (\ref{var_def}), the entire global dynamics can be visualised and analysed easily without resorting to analysis of points or trajectories at infinity. 

We look again to vacuum cosmology in $R^n$ theories to compare the different dynamical systems formulations. The same constraint (\ref{const_mon}) applies to the compact system and using the additional constraints (\ref{friedconstraint}) and (\ref{compactconstraint}), the phase space can be reduced to the 2-dimensional $\bar{Q}$-$\bar{x}$ plane, again without fixing the global spatial curvature. For $R^2$ gravity with a dust fluid ($w=0$) , the dynamical system reduces to 
\begin{eqnarray}
 \frac{d\bar{x}}{d\bar{\tau}}=&& \bar{x}(\bar{Q}+\bar{x})(\bar{x}^2-1),\\
 \frac{d\bar{Q}}{d\bar{\tau}u} =&& \frac{1}{6}\Big(1-\bar{x}^2+\bar{Q}^2(6\bar{x}^2-2)+2\bar{Q}\bar{x}(1+3\bar{x}^2)\Big).
\end{eqnarray}

\begin{table}[h!]
    \centering
\begin{tabular}{|c|c|c|c|}
\hline
    Fixed Point & $(\bar{x}, \bar{Q}, \bar{y}, \bar{K}, \bar{z})$ & Stability & Cosmology \\
    \hline
    $P_{1+}$ & $( 1, 0, 0, 0,0)$ & Unstable & $H=const$\\
    \hline
    $P_{1-}$ & $(-1, 0, 0, 0,0)$ & Stable & $H = const$\\
    \hline
    $P_{2+}$ & $( -1,2, 0,0,0)$ & Unstable & $a\sim t^{1/2}$ \\
    \hline
    $P_{2-}$ & $(1, -2, 0, 0,0)$ & Stable& $a\sim t^{1/2}$\\
     \hline
    $P_{3+}$ & $(0,  \frac{1}{\sqrt{2}}, 1, 0, \frac{1}{2})$ & Stable & $H=const$\\
     \hline
      $P_{3-}$ & $(0, - \frac{1}{\sqrt{2}}, 1, 0, \frac{1}{2})$ & Unstable & $H=const$\\
     \hline
    $P_{4+}$ & $( -\frac{1}{\sqrt{5}}, \frac{1}{\sqrt{5}}, \frac{4}{5}, \frac{3}{5}, \frac{2}{5})$ & Saddle & $a\sim t$ \\
    \hline
    $P_{4-}$ & $(\frac{1}{\sqrt{5}},  -\frac{1}{\sqrt{5}}, \frac{4}{5}, \frac{3}{5}, \frac{2}{5})$ & Saddle &  $a\sim t$\\
    \hline
\end{tabular}
    \caption{Fixed points in the compact phase space of vacuum cosmology in $R^2$ gravity.}
        \label{tab:compactfps}
\end{table}

\begin{figure}[h!]
    \centering
    \includegraphics[scale=0.2]{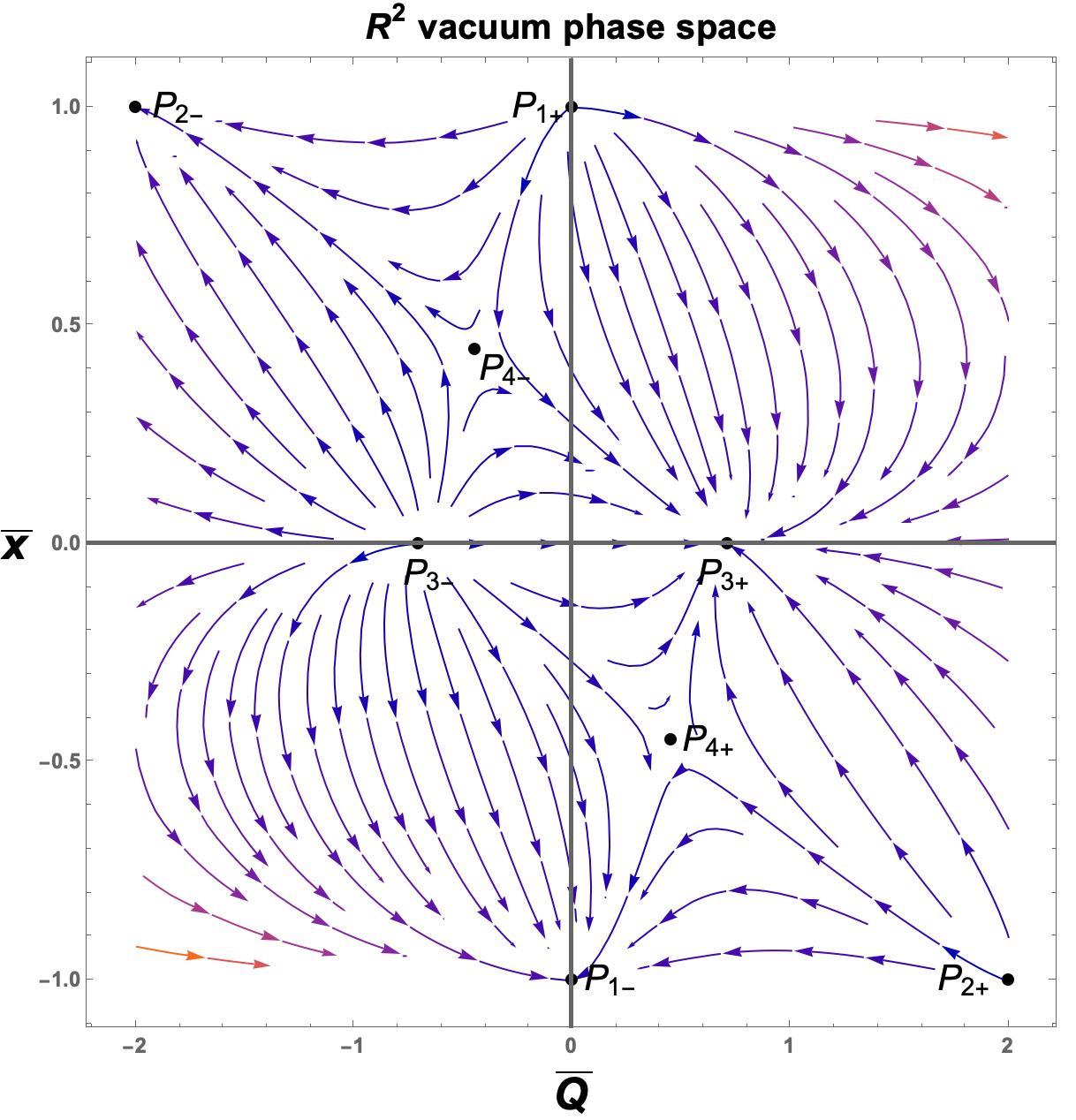}
    \caption{Phase portrait of vacuum solutions of $R^2$ gravity.}
    \label{fig:fig7}
\end{figure}

There are four pairs of fixed points, given in table \ref{tab:compactfps}. Each point has an expanding and collapsing solution, represented by the subscripts $_+$ and $_-$ respectively, save for $P_1$ which represents a static cosmology. The compact phase space is shown in Fig. \ref{fig:fig7}. Unlike in the non-compact formulation, here trajectories representing a number of different cosmological evolutions and can be easily seen. It is clear that cosmological bounces exist for this $f(R)$ model: consider trajectories starting near the static repellor $P_{1+}$, crossing $\bar{Q}=0$ and evolving to the attractor $P_{3+}$. Similarly, trajectories evolving to static solutions ($P_{1-}$ for example) are also possible. This is a novel result for non-spatially flat $R^n$ gravity as previous studies necessarily employ $k =0$ to reduce the phase space dimensionality. 

\begin{figure}[h!]
    \centering
    \includegraphics[scale=0.2]{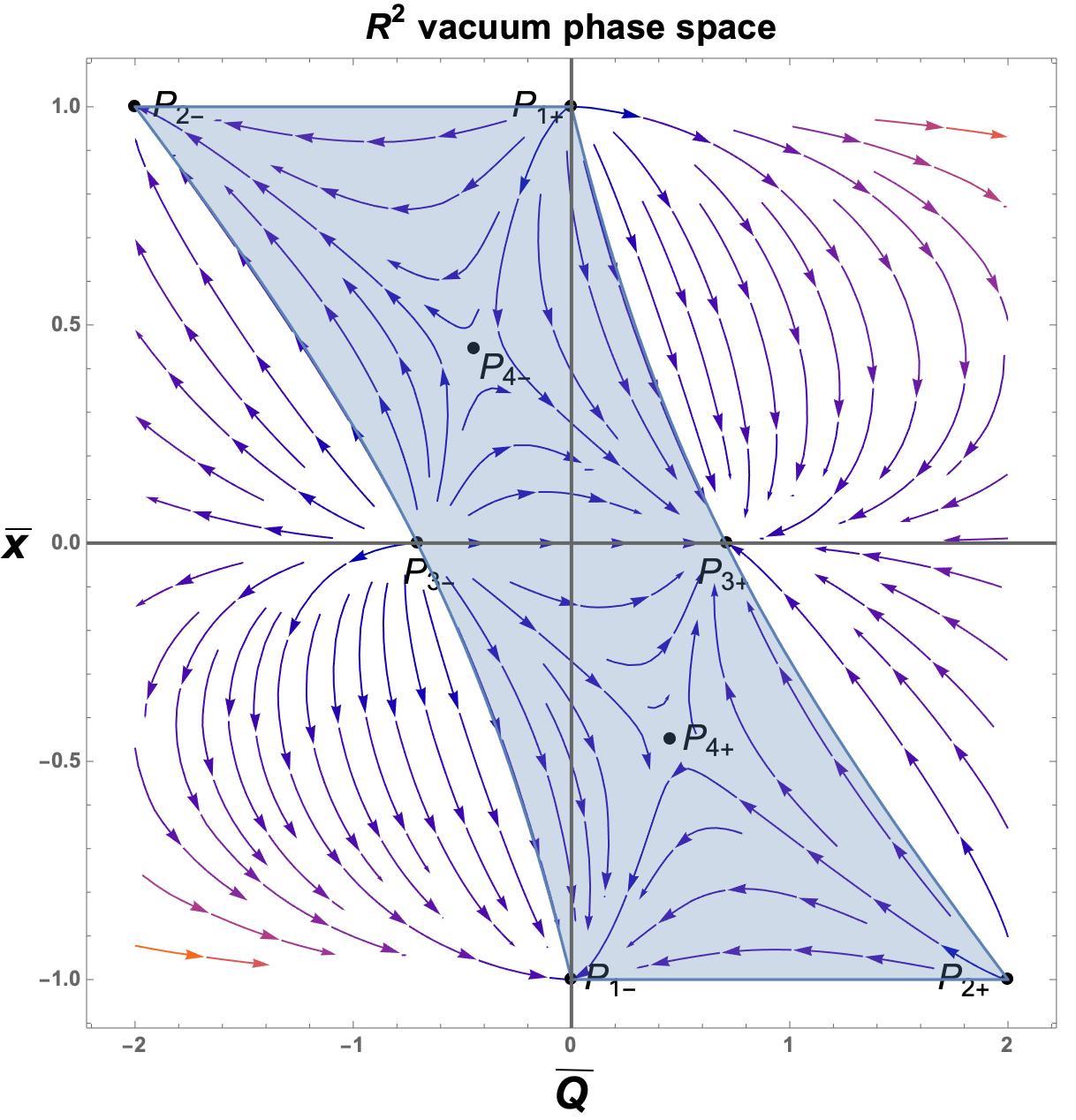}
    \caption{2-dimensional phase portrait of vacuum solutions of $R^2$ gravity, showing the region where viable trajectories in the full 3-dimensional phase space occur.}
    \label{fig:fig8}
\end{figure}

This formulation suffers from the same limitation as the non-compact formalism, that being the necessary invertibility of relation (\ref{yz_f}). In addition, the constraint equations exclude portions of the phase space within the compact region. For example, in the vacuum $R^2$ case, the Friedmann and compactification constraints and the additional constraint (\ref{const_mon}) together give 
\begin{equation}
    \bar{K} = 1 - \frac{(1 - \bar{x}^2)}{2} - (\bar{Q} + \bar{x})^2.
\end{equation}
Enforcing $0\leq \bar{K} \leq 1$ gives restrictions on the viable region of the 2-dimensional $\bar{Q}$-$\bar{x}$ plane such that 
\begin{equation}
    -\bar{x} - \sqrt{\frac{1 + \bar{x}^2}{2}} \leq  
 \bar{Q} \leq -\bar{x} + \sqrt{\frac{1 + \bar{x}^2}{2}}.
\end{equation}
In the 2-dimensional phase portraits it is necessary to plot this region to identify viable trajectories. For this example, the viable region simplifies considerably, however in more complex, non-vacuum cases where the dimensionality of the system cannot be reduced significantly, one may require multiple projections of the phase space - with the additional restrictions - to adequately visualise the global dynamics. This region is shown in Fig. \ref{fig:fig8} where it is now clear that only trajectories in the shaded region are a viable representation of the full 3-dimensional phase space.

\section{Behaviour of cosmological perturbations}

The evolution of cosmological perturbations is dependent on how the background evolves. This is clear from the fact that the coefficients appearing in the evolution equation for linear perturbations consist of background quantities. In fact, the perturbation equations can be written such that the coefficients are expressed completely in terms of the dynamical variables (and possibly the cosmographic parameters). This helps us to find the evolution of a perturbation quantity along different phase trajectories. At the fixed points the perturbation equations simplify to second order differential equations with constant coefficients, which give as solutions two clearly defined perturbation modes. Therefore, a knowledge of the background dynamics as obtained from a phase space analysis can in fact help us conclude something regarding the cosmological perturbations as long as linear cosmological perturbations are considered. Incorporating the study of cosmological perturbations in the phase space picture has merits in several aspects. We discuss three such aspects below.

\subsection{Spectral index for scalar perturbations around a fixed point}

One of the merits is that it allows to connect the qualitative mathematical tool of dynamical system analysis with actual cosmological observables. Let us consider purely scalaron dominated fixed points and introduce the parameters
\begin{equation}
\epsilon_1 = -\frac{\dot{H}}{H^2}, \quad \epsilon_2 = \frac{\dot{F}}{2HF}, \quad \epsilon_3 = \frac{\ddot{F}}{H\dot{F}}.
\end{equation}
These are the so-called slow-roll parameters defined in the context of the inflationary paradigm for early universe, with slow-roll dynamics being specified as $|\epsilon_i|\ll 1$. The three parameters are not actually independent, as the Raychaudhuri equation (\ref{fe2}) can be expressed as
\begin{equation}\label{eps_3}
\epsilon_1 = - \epsilon_2 (1 - \epsilon_3)
\end{equation}
for the spatially flat case. One of the important characteristics related to the cosmological perturbations is the power spectrum, which characterizes how the total amplitude of perturbations is distributed over various it's Fourier modes. This is characterized by a quantity called the spectral index. For a scalaron dominated phase in a spatially flat cosmology, assuming the $\epsilon_i$'s are constant, the scalar spectral index $n_s$ (i.e. the spectral index for scalar cosmological perturbations) can be calculated in terms of the slow-roll parameters \cite{Chiba:2018cmn,Hwang:2001pu}
\begin{equation}
n_s - 1 = 3 - 2\sqrt{\frac{1}{4} + \frac{(1 + \epsilon_1 - \epsilon_2 + \epsilon_3)(2 - \epsilon_2 + \epsilon_3)}{(1 - \epsilon_1)^2}}.
\end{equation}
$n_s=1$ implies a scale-invariant power spectrum, i.e. when the total amplitude of perturbations are evenly distributed over all it's Fourier modes. Slow-roll parameters $\epsilon_1$ and $\epsilon_2$ can be directly related to the dynamical variables
\begin{equation}
\epsilon_1 = 1+q = 2-y+K, \quad\quad \epsilon_2 = \frac{1}{2}x,
\end{equation}
$q$ being the deceleration parameter. This helps us calculate the scalar spectral index for cosmological perturbations at the scalaron dominated fixed points. Since by definition fixed points correspond to solutions for which dynamical variables are constant in time, the slow roll parameters are constant as well. This allows us to derive the following recursion relations that are true for any fixed point
\begin{eqnarray}
&& \dot{H}=-\epsilon_{1}H^2,\,\ddot{H}=2\epsilon_1^{2}H^3,\,\dddot{H}=-6\epsilon_1^{3}H^4,............\,H^{(n)}=(-1)^{n}n!\epsilon_1^{n}H^{n+1},\label{rec_reln_1}
\\
&& \dot{R}=-2\epsilon_{1}HR,\,\ddot{R}=6\epsilon_1^{2}H^{2}R,\,\dddot{R}=-24\epsilon_1^{3}H^{3}R,............\,R^{(n)}=(-1)^{n}(n+1)!\epsilon_1^{n}H^{n}R,\label{rec_reln_2}\nonumber
\\
&&
\\
&& F'=-\frac{\epsilon_2}{\epsilon_1}\frac{F}{R},............\,F^{(n)}=(-1)^n\frac{\epsilon_2}{\epsilon_1}\left(1+\frac{\epsilon_2}{\epsilon_1}\right)\left(2+\frac{\epsilon_2}{\epsilon_1}\right)......\left((n-1)+\frac{\epsilon_2}{\epsilon_1}\right)\frac{F}{R^n}.\label{rec_reln_3}\nonumber
\\
&&
\end{eqnarray}
Recursion relations (\ref{rec_reln_1}) and (\ref{rec_reln_2}) are used to derive recursion relation (\ref{rec_reln_3}).

The values of the slow-roll parameters and the scalar spectral index are listed in table \ref{tab:4} for the fixed points of spatially flat vacuum cosmologies in $R^2$ gravity (see table \ref{tab:1}) and in table \ref{tab:5} for the vacuum fixed points of spatially flat $\Lambda$CDM mimicking $f(R)$ cosmologies (see table \ref{tab:3}). 
\begin{table}[h]
\centering
\begin{tabular}{|c|c|c|}
        \hline
        Fixed point & $(\epsilon_1,\epsilon_2)$ & $n_s$ \\
        \hline
        $P_1$ & $(0,0)$ & $4-\sqrt{1+4(1+\epsilon_3)(2+\epsilon_3)}$ \\
        \hline
        $P_3$ & $\left(2,-\frac{1}{2}\right)$ & $4$ \\
        \hline
    \end{tabular}
    \caption{Slow-roll parameters and the scalar spectral index at the spatially flat vacuum fixed points of $R^2$ gravity.}
    \label{tab:4}
\end{table}
\begin{table}[h]
\centering
\begin{tabular}{|c|c|c|}
        \hline
        Fixed point & $(\epsilon_1,\epsilon_2)$ & $n_s$ \\
        \hline
        $P_1$ & $(0,0)$ & $4-\sqrt{1+4(1+\epsilon_3)(2+\epsilon_3)}$ \\
        \hline
        $P_2$ & $\left(0,\frac{1}{2}\right)$ & $0$ \\
        \hline
        $P_4$ & $\left(\frac{3}{2},\frac{5-\sqrt{73}}{8}\right)$ & $2.772$ \\
        \hline
        $P_5$ & $\left(\frac{3}{2},\frac{5+\sqrt{73}}{8}\right)$ & $-5.772$ \\
        \hline
    \end{tabular}
    \caption{Slow-roll parameters and the scalar spectral index at the spatially flat vacuum fixed points of $\Lambda$CDM mimicking $f(R)$ cosmologies.}
    \label{tab:5}
\end{table}
Notice that $\epsilon_3$ is indeterminable from Eq.(\ref{eps_3}) at De-Sitter fixed points $(\epsilon_1,\epsilon_2)=(0,0)$; therefore we decide to keep the $\epsilon_3$ to be generic in the expression for $n_s$ at such points. If the power spectrum for scalar perturbations around the De-Sitter fixed point is to be scale-invariant, one needs
\begin{equation}
\epsilon_3=0 \Leftrightarrow \epsilon_1=-\epsilon_2,
\end{equation}
which, by virtue of the recursion relation (\ref{rec_reln_3}), translates into the requirement
\begin{equation}
RF'=F.
\end{equation}
It is interesting to note that the above requirement is identically satisfied for all $R$ in case of $R^2$ gravity and never satisfied by any $R^n$ gravity for $n\neq2$. We recover the important result that the $R^2$ gravity is a special case which automatically gives rise to a scale-invariant power spectrum for scalar-perturbations around a De-Sitter phase \cite{DeFelice:2010aj}. If any generic $f(R)$ gravity has a De-Sitter fixed point, then scale-invariance of the power spectrum for scalar perturbations around the point demands that the $f(R)$ theory must be asymptotically $R^2$ at that point.

\subsection{Evolution of matter density contrast near the matter dominated fixed point}

\begin{figure}[h]
\centering
\includegraphics[scale=0.8]{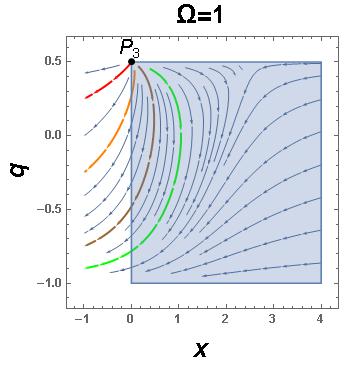}
\caption{\label{fig:6} Projection onto the plane $\Omega=1$ of the phase space of spatially flat $\Lambda$CDM mimicking $f(R)$ cosmologies. Shaded region is where the condition $F'>0$ is met, as given by Eq.(\ref{phys_const}).}
\end{figure}

As another example of how incorporating the study of cosmological perturbations in the phase space picture can be useful, let us consider perturbations in the fluid around the vicinity of the matter dominated fixed point. The evolution equation for the matter density contrast $\delta\equiv\frac{\delta\rho}{\rho}$ is \cite{DeFelice:2010aj}
\begin{equation}\label{ptbn_eqn_1}
\frac{d^2 \delta}{d\tau^2} + (1-q)\frac{d\delta}{d\tau} - \Omega\left(\frac{2x + 3(12\Omega-q-5)\left(\frac{aH}{k}\right)^2}{x + 2(12\Omega-q-5)\left(\frac{aH}{k}\right)^2}\right)\delta = 0.
\end{equation}
The subhorizon modes ($k\gg aH$) evolve through two regimes; a \emph{GR regime} ($0<|x|\ll(aH/k)^2$) and an \emph{$f(R)$ regime} ($0<(aH/k)^2\ll|x|$), with the transition occuring at $|x|\simeq(aH/k)^2$. For an observationally significant $k-$ mode, the faster is the transition from GR regime to $f(R)$ regime, the more is the observational deviation of the $f(R)$ model from the actual $\Lambda$CDM model at the perturbation level. Let us now get back to the 3-dimensional phase space of $\Lambda$CDM mimicking spatially flat $f(R)$ cosmologies and consider it's projection on the $\Omega=1$ plane (Fig.\ref{fig:6}).
No portion of the red trajectory and only a very little portion of the yellow trajectory remain in the shaded region, which represents the projection of the phase space region in which the condition $F'>0$ is met (see Eq.(\ref{phys_const})). Both the brown and green trajectories have significant portion within the shaded region. The green trajectory has a bigger portion within the shaded region, but also deviates from $x=0$ (GR limit) faster, as compared to the brown trajectory. For an observationally relevant perturbation $k$-mode, the transition from GR regime to $f(R)$ regime is faster in the green trajectory than the brown one. Therefore, whereas the $f(R)$ cosmology represented by the green trajectory seems to be a better choice when removing the tachyonic instability is in mind, it also gives rise to more observational deviation from the actual $\Lambda$CDM model at the perturbation level.

\subsection{Covariant approach to cosmological perturbation using the 1+3 decomposition}

The usual approach to cosmological perturbations relies on the so-called ``slicing and threading'' \cite{Mukhanov:1990me}, which ruins the covariance of the picture. This raises the issue of gauge-dependence of cosmological perturbations, which is tackled either by choosing a particular gauge or by cleverly defining gauge-invariant perturbation variables \cite{osti_5003202}. An alternative covariant approach to cosmological perturbations was first suggested by Hawking \cite{osti_4529555} and later developed by others \cite{Ellis:1989ju,Bruni:1992dg,Ellis:1989jt,Alho:2013vva}. This approach is based on the so-called 1+3 covariant decomposition approach, in which one writes down a generic form of the field equations in terms of a set of covariant quantities defined locally with respect to an arbitrary timelike observer \cite{Ellis:1998ct}. Homogeneous and isotropic FLRW background corresponds to the vanishing of some of these covariant quantities, which, when assumed to be up to first order of smallness, correspond to linear perturbations. Because all the quantities are by definition covariant, including the perturbation quantities, they are also manifestly gauge-invariant. This approach has been successfully implemented to $f(R)$ cosmology \cite{Carloni:2007yv,Carloni:2008jy,Ananda:2008tx,Abebe:2011ry}. There are five scalar perturbation variables
\begin{equation}
\Delta_m \equiv \frac{a^2}{\rho_m}\tilde{\nabla}^2\rho_m, \quad 
Z \equiv 3a^2 \tilde{\nabla}^2 H, \quad
C \equiv a^4 \tilde{\nabla}^2 \tilde{R}, \quad
\mathcal{R} \equiv a^2 \tilde{\nabla}^2 R, \quad
\Re \equiv a^2 \tilde{\nabla}\dot{R}.
\end{equation}
where $\tilde{\nabla}$ is the total orthogonal projection of the covariant derivative on the hypersurface orthogonal to the worldline of the timelike observer. They are related by the constraint
\begin{equation}
    \frac{C}{a^2} + \left(4H + 2\frac{\dot{f'}}{f'}\right) Z - 2 \frac{\rho_m}{f'}\Delta_m + \left[6H\frac{\dot{f''}}{f'} - \frac{f''}{f^{\prime 2}}(f - 2\rho_m + 6H\dot{f'}) + 2\frac{f''}{f'}\frac{l^2}{a^2}\right]\mathcal{R} + 6H\frac{f''}{f'}\Re = 0.
\end{equation}    
Existence of the constraint allows us to eliminate one of the perturbation variables. If we choose to eliminate $C$, the dynamical equations corresponding to the other quantities are \cite{Carloni:2008jy}
\begin{eqnarray}
&& \dot{\Delta}_{m} = 3wH\Delta_{m} - (1+w)Z,
\\
&& \dot{Z} = \left(\frac{\dot{f'}}{f'} - 2H\right)Z + \left[\frac{(w-1)(3w+2)}{2(w+1)}\frac{\rho_m}{f'} + \frac{18wH^2 + 3w(\rho_R + 3p_R)}{6(w+1)}\right] \Delta_{m} + 3\frac{Hf''}{f'} \Re \nonumber
\\
&& \hspace{10mm} + \left[\frac{1}{2} - \frac{f''}{f'}\frac{l^2}{a^2} - \frac{1}{2}\frac{ff''}{f^{\prime 2}} - \frac{f''\rho_m}{f^{\prime 2}} + 3\dot{R}H\left(\frac{f''}{f'}\right)^{2} + 3\dot{R}H\frac{f'''}{f'}\right]\mathcal{R},
\\
&& \dot{\mathcal{R}} = \Re - \frac{w}{w+1}\dot{R}\Delta_{m},
\\
&& \dot{\Re} = - \left(3H + 2\frac{\dot{f''}}{f''}\right)\Re - \dot{R}Z - \left[\frac{(3w-1)}{3}\frac{\rho_m}{f''} + \frac{w}{3(w+1)}\ddot{R}\right]\Delta_{m}\nonumber
\\
&& + \left[\frac{k^2}{a^{2}} - \left(\frac{1}{3}\frac{f'}{f''} + \frac{\dot{R}\dot{f'''}}{f'} + 3H\frac{\dot{f''}}{f''} + \ddot{R}\frac{f'''}{f''} -\frac{R}{3}\right)\right] \mathcal{R}.
\end{eqnarray}
Instead of the four first order coupled differential equations above, the system can be represented in a neat way as two second order coupled differential equations. It is possible to write the perturbation equations with coefficients expressed in terms of the dynamical variables by redefining the curvature perturbation variable as
\begin{equation}
    \mathcal{R} = a^2 \tilde{\nabla}^2 R \rightarrow \mathcal{R} = a^2 \tilde{\nabla}^2 \ln[f'(R)].
\end{equation}
Note that at the GR limit $f'(R)\rightarrow1$, $\mathcal{R}\rightarrow0$, so that this redefined curvature perturbation variable encapsulates the deviation of the underlying theory from GR. In terms of this redefined curvature perturbation variable and following Ref.\cite{Carloni:2008jy} we can write
\begin{eqnarray}
&& \frac{d^2 \Delta_m}{d\tau^2} + \mathcal{A}\frac{d\Delta_m}{d\tau} + \mathcal{B}\Delta_m + \mathcal{C}\mathcal{R} + \mathcal{D}\frac{d\mathcal{R}}{d\tau} = 0,
\label{ptbn_eqn_21}\\
&& \frac{d^2 \mathcal{R}}{d\tau^2} + \mathcal{E}\frac{d\mathcal{R}}{d\tau} + \mathcal{F}\mathcal{R} + \mathcal{G}\Delta_m + \mathcal{H}\frac{d\Delta_m}{d\tau} = 0, \label{ptbn_eqn_22}
\end{eqnarray}
where $\epsilon\equiv\frac{H}{|H|}$ and the coefficients are as follows
\begin{eqnarray}
&& \mathcal{A} = \epsilon(1-3w+z -\Omega),\\
&& \mathcal{B} = -3(2wz - 3wK+(1-w)\Omega),\\
&& \mathcal{C} = -3(1+w)(z- \Gamma y -3K-\Omega),\\
&& \mathcal{D} = 3\epsilon(1+w),\\
&& \mathcal{E} = -\epsilon(3x-3y+2z-2\Omega+1),\\
&& \mathcal{F} = 4z- 2\Gamma y - K-(1-3w)\Omega,\\
&& \mathcal{G} = \frac{4w(y-2z)-(1-4w+3w^2)\Omega}{1 + w},\\
&& \mathcal{H} = \epsilon\left(\frac{1-w}{1+w}\right)(1+x-y+z-\Omega).
\end{eqnarray}

Using Eq.(\ref{var_reln}) we can also write the perturbation equations with coefficients expressed in terms of the compact dynamical variables as follows
\begin{eqnarray}
&& \frac{d^2 \Delta_m}{d\tau^2} + \bar{\mathcal{A}}\frac{d\Delta_m}{d\tau} + \bar{\mathcal{B}}\Delta_m + \bar{\mathcal{C}}\mathcal{R} + \bar{\mathcal{D}}\frac{d\mathcal{R}}{d\tau} = 0,
\\
&& \frac{d^2 \mathcal{R}}{d\tau^2} + \bar{\mathcal{E}}\frac{d\mathcal{R}}{d\tau} + \bar{\mathcal{F}}\mathcal{R} + \bar{\mathcal{G}}\Delta_m + \bar{\mathcal{H}}\frac{d\Delta_m}{d\tau} = 0, 
\end{eqnarray}
where
\begin{eqnarray}
&& \bar{\mathcal{A}} = \epsilon[(1-3w)\bar{Q}^2 + \bar{z} - \bar{\Omega}],\\
&& \bar{\mathcal{B}} = -3(2w\bar{z} - 3w\bar{K} + (1-w)\bar{\Omega}),\\
&& \bar{\mathcal{C}} = -3(1+w)(\bar{z} - \Gamma \bar{y} - 3\bar{K} - \bar{\Omega}),\\
&& \bar{\mathcal{D}} = 3\epsilon(1+w)\bar{Q}^2,\\
&& \bar{\mathcal{E}} = -\epsilon(6\bar{x} \bar{Q} - 3\bar{y} + 2\bar{z} - 2\bar{\Omega} + 1),\\
&& \bar{\mathcal{F}} = 4\bar{z} - 2\Gamma \bar{y} - 9\bar{K} - (1-3w)\bar{\Omega},\\
&& \bar{\mathcal{G}} = \frac{4w(\bar{y} - 2\bar{z}) - (1-4w+3w^2)\bar{\Omega}}{1+w},\\
&& \bar{\mathcal{H}} = \epsilon\left(\frac{1-w}{1+w}\right)(\bar{Q}^2 + 2\bar{x} \bar{Q} - \bar{y} + \bar{z} - \bar{\Omega}).
\end{eqnarray}
We further note that 
\begin{equation}
    \frac{d}{d\tau} = \frac{\epsilon}{\bar{Q}}\frac{d}{d\bar{\tau}}, \qquad \frac{d^2}{d\tau^2} = \frac{\epsilon}{\bar{Q}}\frac{d^2}{d\bar{\tau}^2} - \frac{\epsilon}{\bar{Q}^2}\frac{d\bar{Q}}{d\bar{\tau}}\frac{d}{d\bar{\tau}},
\end{equation}
so that we can write the perturbation equations as differential equations with respect to $\bar{\tau}$
\begin{eqnarray}
&& \frac{d^2\Delta_m}{d\bar{\tau}^2} + \left(\bar{\mathcal{A}} - \frac{d\ln \bar{Q}}{d\bar{\tau}}\right)\frac{d\Delta_m}{d\bar{\tau}} + \frac{\bar{Q}}{\epsilon}\bar{\mathcal{B}} \Delta_m \nonumber \\
&& \hspace{40mm} + \frac{\bar{Q}}{\epsilon}\bar{\mathcal{C}} \mathcal{R} + \bar{\mathcal{D}} \frac{d\mathcal{R}}{d\bar{\tau}} = 0,
\label{ptbn_eqn_31}\\
&& \frac{d^2\mathcal{R}}{d\bar{\tau}^2} + \left(\bar{\mathcal{E}} - \frac{d\ln \bar{Q}}{d\bar{\tau}}\right)\frac{d\mathcal{R}}{d\bar{\tau}} + \frac{\bar{Q}}{\epsilon}\bar{\mathcal{F}} \mathcal{R} \nonumber \\
&& \hspace{40mm} + \frac{\bar{Q}}{\epsilon}\bar{\mathcal{G}} \Delta_m + \bar{\mathcal{H}} \frac{d\Delta_m}{d\bar{\tau}} = 0. \label{ptbn_eqn_32}
\end{eqnarray}
perturbation equations (\ref{ptbn_eqn_21},\ref{ptbn_eqn_22}) or (\ref{ptbn_eqn_31},\ref{ptbn_eqn_32}) enables us to analyze the behavior of cosmological perturbations corresponding to a background evolution. A particular background evolution, \emph{i.e.} a particular phase trajectory is parametrically given by $(x(\tau),y(\tau),z(\tau),\Omega(\tau),K(\tau))$ or $(\bar{Q}(\bar{\tau}),\bar{x}(\bar{\tau}),\bar{y}(\bar{\tau}),\bar{z}(\bar{\tau}),\bar{\Omega}(\bar{\tau}),\bar{K}(\bar{\tau}))$, which specifies the coefficients of the perturbation equations (\ref{ptbn_eqn_21},\ref{ptbn_eqn_22}) or (\ref{ptbn_eqn_31},\ref{ptbn_eqn_32}). These equations can be solved numerically to determine the behaviour of the two scalar perturbation modes $\Delta_m$ and $\mathcal{R}$. For fixed points corresponding to a non-static cosmology ($\bar{Q}\neq0$), Eqs.(\ref{ptbn_eqn_31},\ref{ptbn_eqn_32}) simplifies to
\begin{eqnarray}
&& \frac{d^2\Delta_m}{d\bar{\tau}^2} + \bar{\mathcal{A}} \frac{d\Delta_m}{d\bar{\tau}} + \frac{\bar{Q}}{\epsilon}\bar{\mathcal{B}} \Delta_m + \frac{\bar{Q}}{\epsilon}\bar{\mathcal{C}} \mathcal{R} + \bar{\mathcal{D}} \frac{d\mathcal{R}}{d\bar{\tau}} = 0,
\\
&& \frac{d^2\mathcal{R}}{d\bar{\tau}^2} + \bar{\mathcal{E}} \frac{d\mathcal{R}}{d\bar{\tau}} + \frac{\bar{Q}}{\epsilon}\bar{\mathcal{F}} \mathcal{R} + \frac{\bar{Q}}{\epsilon}\bar{\mathcal{G}} \Delta_m + \bar{\mathcal{H}} \frac{d\Delta_m}{d\bar{\tau}} = 0.\nonumber\\
&& 
\end{eqnarray}

\section{Discussion}
\label{sec:disc}

There is no unique way to formulate dynamical system for $f(R)$ gravity theories (or any theory whatsoever). The only requirement is that we should be able to obtain an autonomous system. Over the past ten years there has been a considerable amount of work done on exploring the cosmological dynamics of $f(R)$ gravity. It is worth mentioning that this review has mostly focused on using variables that lead to a non-compact phase space. As shown in Sec. \ref{sec:compact}, one can define variables which are positive definite throughout phase space, compactifying the phase space and allowing for a detailed analysis of the asymptotics and other features such as the existence (and stability) of Einstein Static points \cite{Goswami:2008fs} and possible cyclic behaviour \cite{Abdelwahab:2011dk}. Finally, all the methods discussed here can be easily extended to Bianchi models \cite{Goheer:2007wu,Chakraborty:2018bxh}. In the case of Bianchi I cosmologies, it was shown that the existence of an additional scalar degree of freedom leads to the possibility of both past and future isotropic states \cite{Leach:2007ss}. In conclusion, the dynamics of $f(R)$ gravity is an extremely rich area of study and there remains still much interesting work to be done.

\bibliographystyle{unsrt}
\bibliography{citation}

\end{document}